\documentclass[twocolumn,prb]{revtex4}
\usepackage{graphicx}    
\usepackage{dcolumn}     
\usepackage{bm}          
\newcommand{\Vec}[1]{\mbox{\boldmath$#1$}}
\begin{document}
\title{Spin fluctuations and unconventional pairing 
in KFe$_2$As$_2$}
\author{Katsuhiro Suzuki$^{1,3}$, Hidetomo Usui$^2$, and Kazuhiko Kuroki$^{1,3}$ }
\affiliation{$^1$Department of Engineering Science, 
The University of Electro-Communications, Chofu, Tokyo 182-8585, Japan}
\affiliation{$^2$Department of Applied Physics and Chemistry, 
The University of Electro-Communications, Chofu, Tokyo 182-8585, Japan}
\affiliation{$^3$JST, TRIP, Chofu, Tokyo 182-8585, Japan}
\date{\today}
\begin{abstract}
We study the relation between the spin fluctuation and 
superconductivity in a heavily hole doped end material 
KFe$_2$As$_2$. We construct a five orbital model 
by approximately unfolding the Brillouin zone of the 
three-dimensional ten-orbital model obtained from first-principles calculation.
By applying the random-phase-approximation, we obtain the 
spin susceptibility and solve the linearized Eliashberg equation.
The incommensurate spin fluctuation observed experimentally 
is understood as originating from interband interactions, 
where the multiorbital nature of the band structure 
results in an electron-hole asymmetry of the incommensurability 
in the whole iron-based superconductor family.
As for superconductivity, $s$-wave and $d$-wave pairings are 
found to be in close competition, where the sign change in the 
gap function in the former is driven by 
the incommensurate spin fluctuations.
We raise several possible explanations for the 
nodes in the superconducting gap of KFe$_2$As$_2$ observed 
experimentally.
\end{abstract}
\maketitle

\section{Introduction}
A number of interesting features of the iron-based superconductors have 
attracted much attention.
Among them is the material dependence of the superconductivity. 
Not only the transition temperature $T_c$, but also the 
form of the superconducting gap seems to be material dependent.
For instance, although a number of experiments suggest that the 
gap is fully open on the Fermi surface on some of the 
arsenides, there are several experiments suggesting the presence of 
nodes in the superconducting gap LaFePO\cite{Fletcher,Hicks,Yamashita}, 
BaFe$_2$(As$_{1-x}$P$_x$)$_2$ \cite{Hashimoto,Ishida,Yamashita2}, 
or in LiFeP\cite{HashimotoLi111}.
Also in the electron-doped 122 systems, 
Ba(Fe$_{1-x}$M$_x$)$_2$As$_2$(M=Co,Ni), the presence of gap nodes 
has been suggested.\cite{Martin,Mazin122} 

For LaFePO, theoretical studies suggest the presence of nodes on
the electron Fermi surface around the wave vector 
$(\pi,0)/(0,\pi)$ in the unfolded Brillouin zone.
\cite{Kuroki_prb,DHLeeP,IkedaFLEX2,Kariyadoreal,ThomaleLaFePO}
From this viewpoint, the heavily hole doped end material KFe$_2$As$_2$ is 
of special interest because in this material, the electron pockets  
present in most of the iron-based superconductors are known to 
be absent due to the low position of the Fermi energy.\cite{Sato,Terashima}
Nonetheless, the material becomes superconducting, and interestingly, 
several  experiments suggest the presence of nodes in the gap.
\cite{Fukazawa,Dong,Zhang,Matsuda2} 
Since there are no appreciable electron pockets in KFe$_2$As$_2$, 
the nodes are likely to be on the hole Fermi surface.
Theoretically, a possibility of $d$-wave pairing has been proposed 
recently.\cite{ThomaleK122,Maiti}

Another interesting observation in KFe$_2$As$_2$ is the 
spin fluctuations found in neutron-scattering experiments.
In Ref. \onlinecite{LeeK122}, incommensurate spin fluctuation has been observed, 
and we have theoretically provided its explanation in terms of the 
multiorbital nature of the Fermi surface.

Given this background, the aim of the present study is to 
understand the relation between the spin fluctuations and the 
superconducting gap of KFe$_2$As$_2$.
We first construct a ten-orbital model for 
KFe$_2$As$_2$ and also obtain a five-orbital model by 
approximately unfolding the Brillouin zone.
We adopt three-dimensional models in order to take into 
account the possible variation of the superconducting gap 
along the $k_z$ direction.
We apply random phase approximation to these models and 
compare the calculation results at  high temperature, 
which gives reasonable agreement, 
although the Brillouin zone unfolding is not exact.
Then, we go down to low temperature using the five-orbital model.
We first summarize our understanding 
on the origin of the incommensurate spin fluctuations\cite{LeeK122} 
and also analyze its strength when the material composition is 
varied from BaFe$_2$As$_2$ to KFe$_2$As$_2$. 
We finally discuss the 
superconducting gap when Cooper pairing is mediated by spin
fluctuations. An analysis based on the linearized 
Eliashberg equation shows that 
$s$-wave and $d$-wave pairings are found to be in close 
competition, and the former 
is found to have essentially the $s\pm$-wave form 
in the sense that the gap function around the wave vector 
$(0,0)$ (where the hole Fermi surfaces exist) and around 
$(\pi,0)/(0,\pi)$ (where the electron Fermi surface would have 
existed in a less doped material) has the opposite sign.
This sign change is driven by the incommensurate spin fluctuation, 
which is due to the interband interaction between states around 
(0,0) and $(\pi,0)/(0,\pi)$. Since the electron Fermi pockets are
absent, the latter states are somewhat away from the 
Fermi level, so that this interband sign change  of the gap function 
should not be detected experimentally. 
As for the possible explanation of the nodes of the gap observed 
experimentally, we raise several 
possibilities concerning the hole Fermi surface.

\section{Band structure and Fermi surface} 

\begin{figure}[b]
\begin{center}
\includegraphics[width=7.2cm,clip]{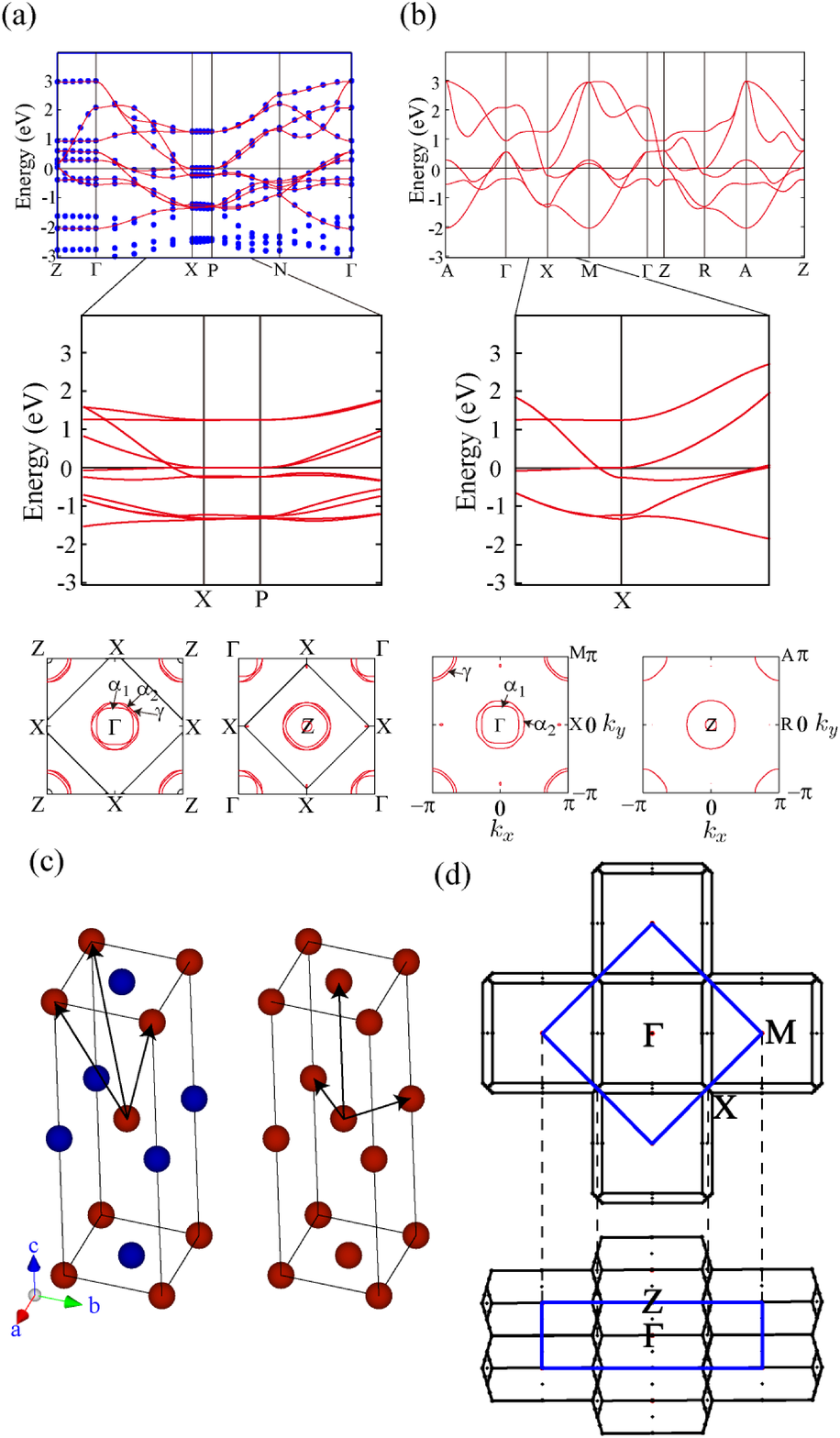}
\end{center}
\caption{(color online) Band structure and Fermi surfaces of 
(a) the ten-orbital model in the original Brillouin zone of the 
BCT lattice structure and (b) the five-orbital model of KFe$_2$As$_2$ 
presented in the unfolded Brillouin zone of the ST lattice structure. 
The dots in the left are the original first-principles band calculation, 
while the solid lines are the tight-binding model band dispersion 
obtained by using maximally localized Wannier orbitals. 
The horizontal cuts of the Fermi surface at $k_z=0$ and $k_z=\pi$ are presented 
in the bottom panels of (a) and (b).  
Translating the left (right) panel of the five-orbital Fermi surface 
(b) by $(\pi,\pi,\pi)$ and superposing it to the right (left) corresponds to the 
ten-orbital Fermi surface shown in the right (left).
(c) The BCT and the ST lattice structure. The former (latter)  unit 
cell with two (one) iron(s) is adopted for the ten (five)-orbital model.
The figure has been produced using VESTA\cite{VESTA}.
(d) The Brillouin zone correspondence between the ten-orbital (black line) 
and five-orbital (blue line) models. The Z point of the second Brillouin zone 
of the BCT structure coincides with the M point of the unfolded Brillouin zone.}
\label{fig1}
\end{figure}

\begin{figure}[b]
\begin{center}
\includegraphics[width=8cm,clip]{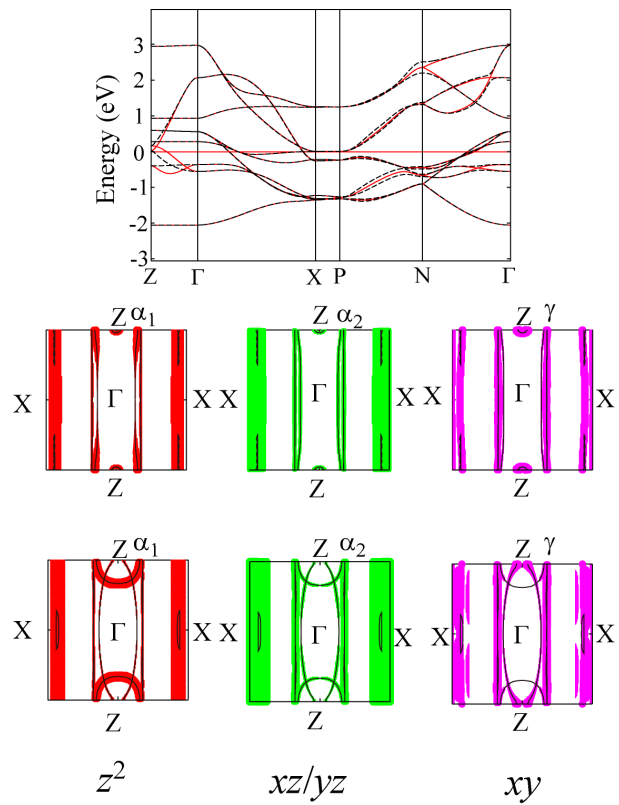}
\end{center}
\caption{(Color online) (Top) The band structure of the five-orbital model 
of KFe$_2$As$_2$ refolded into the original Brillouin zone (solid red), 
and the band structure of the ten-orbital model (dashed black). 
(Bottom panels) Horizontal cuts of the Fermi surface for 
the ten-orbital (upper) and five-orbital (lower) models. 
The thickness represents the $z^2$, $xz/yz$ and $xy$ orbital characters.
}
\label{theory}
\end{figure}

First we perform a first-principles band calculation for 
KFe$_2$As$_2$ adopting the experimentally determined lattice 
structure\cite{struct} and using the QUANTUM ESPRESSO package.\cite{pwscf} 
Then we exploit maximally localized Wannier orbitals\cite{MaxLoc} 
to obtain the ten-orbital model, whose band structure and the 
Fermi surface are shown in Fig. \ref{fig1} (a). 
There are ten $3d$ bands because there are two iron atoms per 
unit cell in the body-centered tetragonal (BCT) lattice structure. 
Around the X point, the electron Fermi surfaces, which are present in
iron pnictides with fewer holes,  are absent. Nonetheless, the band that 
gives rise to the electron Fermi surface for smaller number of holes sits 
very close to the Fermi level. Three cylindrical hole Fermi surfaces are 
present around the $\Gamma$ point, which we name $\alpha_1$, $\gamma$, and 
$\alpha_2$ from the inner one to the outer. 
Also around the Z point there is a tiny hole Fermi pocket.

For 122 systems, the Brillouin zone  unfolding procedure\cite{Kuroki2008} 
that adopts the reduced unit cell (with one iron) cannot be done 
strictly,\cite{Mazin122} namely, the hopping integrals that go out from and 
come into Fe1 and Fe2 in Fig. \ref{fig1}(c) are somewhat different. 
Nonetheless, if we adopt a reduced unit cell by neglecting the difference, 
we end up with an approximate five-orbital model.\cite{GraserComment}
We show the band structure and the Fermi surface of this model in Fig. \ref{fig1} (b) 
in the ``unfolded'' Brillouin zone of the simple tetragonal (ST) lattice structure. 
The correspondence between the unfolded ST and the folded BCT Brillouin zone 
is shown in Fig. \ref{fig1}(d).
As seen from this correspondence, the band structure of the five-orbital model 
can be refolded into the original BCT Brillouin zone by translating the bands 
in the second BCT Brillouin zone by a wave vector ($\pi$,$\pi$,$\pi$). 
The comparison of the band structure between the original ten-orbital model and the 
five-orbital model in the refolded Brillouin zone is shown in Fig. \ref{theory}. 
They coincide with each other in most of the portions, 
but there is some discrepancy along $\Gamma$-Z.

In the middle panels of Fig. \ref{theory}, we show the vertical cuts of the 
Fermi surface, where the thickness represents the strength of the $z^2$, $xz/yz$, 
or $xy=(X^2-Y^2)$ orbital character.
All of the hole Fermi surfaces are very cylindrical, 
so that the two-dimensionality is strong. 
However, if we look at the orbital character, $\alpha_2$ and $\gamma$ 
Fermi surfaces have strong $xz/yz$ and $xy$ character, respectively, 
regardless of $k_z$, while the $\alpha_1$ Fermi surface is mainly 
composed of $z^{2}$ orbital around the Z point, while the $xz/yz$ character 
becomes stronger around the $\Gamma$ point. 
In this sense, $\alpha_1$ can be considered as less two-dimensional 
compared to the other two-hole Fermi surfaces.

In the bottom panels of Fig. \ref{theory}, we show similar vertical cuts of 
the Fermi surface for the five-orbital model, refolded into the original Brillouin zone.
The result in most portions is close to that obtained for the ten-orbital model, 
but the $\alpha_1$ hole Fermi surface, which is a cylinder in the latter, 
splits into a three-dimensional pocket around the Z point and a strongly warped cylinder.
This split occurs between portions of the Fermi surface having strong $z^2$ and $xz/yz$ character.
Although this may seem like a large difference, the effect of this splitting is not so large 
as far as the quantities we focus on are concerned, as we shall see later.

\section{Many-Body Hamiltonian and Random-Phase Approximation}
For the many-body part of the Hamiltonian, we consider the 
intraorbital $U$, the interorbital $U'$, the Hund's coupling $J$, 
and the pair hopping interaction $J'$. We consider orbital-dependent 
interactions,\cite{Miyake} and the interaction part of the  Hamiltonian is given as 
\begin{eqnarray}
\mathcal{H'}=\sum_{i}\left(\sum_{\mu}U_{\mu}n_{i\mu\uparrow}
n_{i\mu\downarrow}+\sum_{\mu>\nu}\sum_{\sigma,\sigma'}U'_{\mu\nu}
n_{i\mu\sigma}n_{i\mu\sigma'}\right.\nonumber\\
\left.-\sum_{\mu\neq\nu}J_{\mu\nu}S_{i\mu}\cdot S_{i\nu}
+\sum_{\mu\neq\nu}J'_{\mu\nu}c^{\dagger}_{i\mu\uparrow}
c^{\dagger}_{i\mu\downarrow}c_{i\nu\uparrow}c_{i\nu\downarrow}
\right), 
\label{eq1}
\end{eqnarray}
where $i$ denotes the sites and $\mu$, $\nu$ stand for the orbitals.
We apply random-phase approximation (RPA) to this model and obtain 
the spin and charge susceptibility matrices. Namely, using the bare 
Green's function $G_{lm}(k)$ $(k=(\Vec{k},i\omega_n))$ in the orbital 
representation, the irreducible susceptibility matrix is given as 
\begin{equation} 
\chi^0_{l_1,l_2,l_3,l_4}(q) =\sum_k G_{l_1l_3}(k+q)G_{l_4l_2}(k)
\end{equation}
The spin and the charge (orbital) susceptibility matrices are obtained as 
\begin{equation}
\chi_s(q)=\frac{\chi^0(q)}{1-S\chi^0(q)}
\end{equation}
\begin{equation}
\chi_c(q)=\frac{\chi^0(q)}{1+C\chi^0(q)}, 
\end{equation}
where $S$ and $C$ are interaction vertex matrices.
We refer to the largest eigenvalue of the matrices $S\chi^0(q)$ and $\chi_s(q)$ 
for $i \omega_n=0$ as the Stoner factor and the spin susceptibility, respectively.
The magnetic instability is signaled by the Stoner factor exceeding unity. 
In the RPA calculation (where the self-energy correction is neglected), 
adopting the interaction values evaluated by first-principles 
calculation\cite{Miyake} easily results in a magnetic instability even 
at high temperature, so we multiply all the electron-electron interaction 
by a constant factor $f$\cite{Kuroki_prb}.

To analyze superconductivity, the Green's function and 
the effective singlet pairing interaction, 
\begin{equation}
V^s(q)=\frac{3}{2}S\chi^s(q)S-\frac{1}{2}C\chi^c(q)C+\frac{1}{2}(S+C),
\end{equation}
are plugged into the linearized Eliashberg equation, 
\begin{eqnarray}
\lambda \phi_{l_1l_4}(k)&=&-\frac{T}{N}\sum_q
\sum_{l_2l_3l_5l_6}V_{l_1 l_2 l_3 l_4}(q)G_{l_2l_5}(k-q)\nonumber \\
&&\times\phi_{l_5l_6}(k-q)G_{l_3l_6}(q-k)
\end{eqnarray}
The eigenvalue $\lambda$ increases upon lowering the temperature 
and reaches unity at $T_c$. 
Instead of going down to $T_c$, which will be a tedious calculation, 
we perform the calculation at a fixed temperature and obtain the 
eigenvalue and the eigenvector for $s$-wave and $d$-wave pairing symmetries. 
The comparison of the eigenvalue at a fixed (low) temperature for different 
pairing symmetries tells us which pairing state is actually realized in the system.
The eigenfunction of the Eliashberg equation is referred to 
as the superconducting gap function.
\section{Comparison between ten- and five-orbital models}
\label{1025sec}
\begin{figure}[t]
\begin{center}
\includegraphics[width=7.5cm,clip]{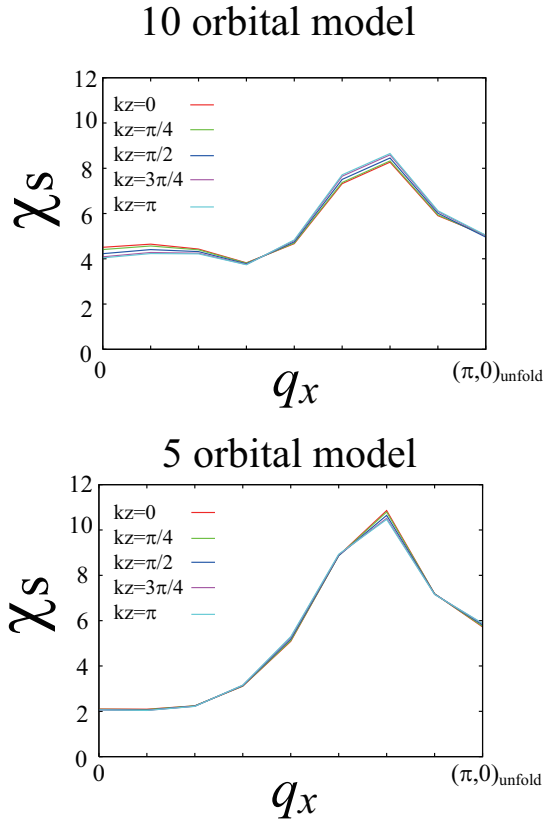}
\end{center}
\caption{(Color online) Comparison of the spin susceptibility 
between the ten-orbital (top) and five-orbital (bottom) models 
along the wave vector $(0,0)$ to ($\pi$,0). 
$T=0.07$eV with $16\times 16\times 16$ $k$-point meshes.}
\label{1025chi}
\end{figure}
We first compare the five- and ten-orbital models 
at a high temperature of $T=0.07$eV.
We take $16\times 16\times 16$ $k$-point meshes and  128 Matsubara 
frequencies. The interaction reducing factor is $f=0.53$.
The Stoner factor calculated for this parameter set is $\alpha_S=0.94$ and 0.95 
for the ten- and five-orbital models, respectively, in reasonable agreement. 
In Fig. \ref{1025chi}, we show the spin susceptibility along 
the wave vector (0,0) to $(\pi,0)$ in the unfolded Brillouin zone. 
The slight difference in the Stoner factor results in a difference 
in the peak value of the spin susceptibility, but the $q_x$ position at which the 
spin susceptibility is maximized is the same between the two models. 
It can also be seen from this figure that the three-dimensionality of 
the spin fluctuations is very small.

\begin{figure}[htbp]
\begin{center}
\includegraphics[width=5.5cm,clip]{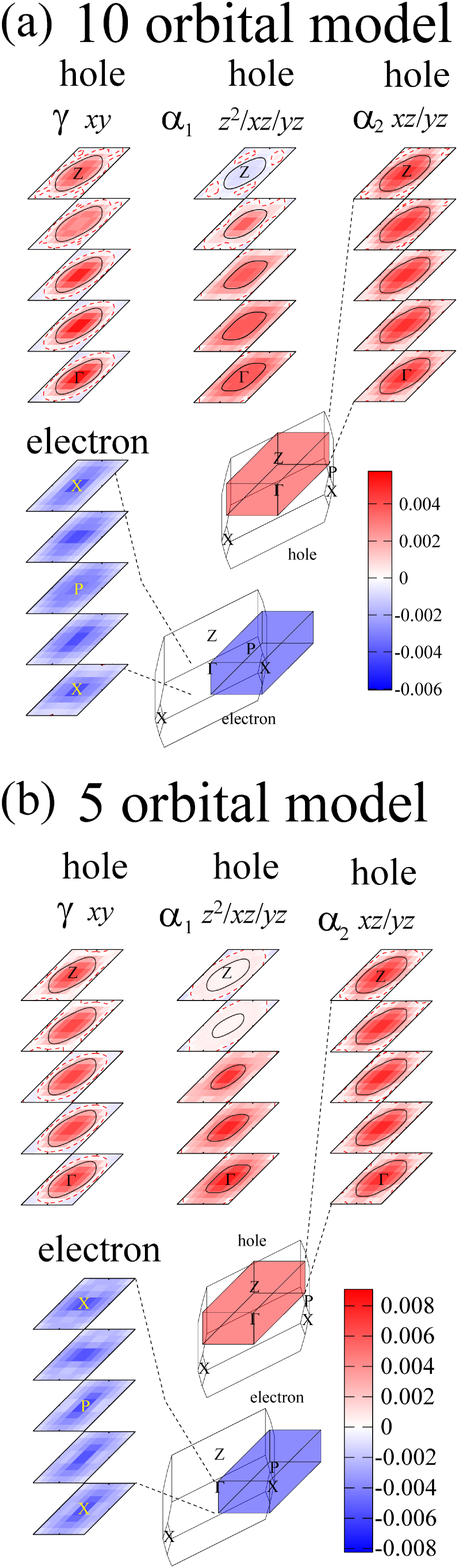}
\end{center}
\caption{(Color online) Comparison of the $s\pm$ gap function
between (a) the ten-orbital and (b) the five-orbital models at 
$T=0.07$eV with $16\times 16\times 16$ $k$-point meshes. 
Contour plots are shown for five horizontal cuts around 
$\Gamma$-Z and X-P-X in the original Brillouin zone of the BCT lattice. 
The five-orbital results, obtained originally in the unfolded 
Brillouin zone, are refolded into this Brillouin zone.
Solid black lines are the Fermi surface and the 
red dashed lines are the nodal lines of the gap.
The Brillouin zone is shown along with the regions 
where the gap is presented.}
\label{1025gap}
\end{figure}

In Fig.\ref{1025gap}, we compare the superconducting gap for the 
two models in the folded Brillouin zone. 
For the five-orbital model, the results are obtained first in the 
unfolded Brillouin zone, and then those in the second Brillouin zone of the 
original BCT lattice structure is moved to the first Brillouin zone.
Here again they have similar structures for most of the 
$k_z$ planes given here. 
The eigenvalue of the Eliashberg equation is 0.66 and 0.76 for $s$-wave, 
and 0.51 and 0.60 for $d$-wave for the ten- and five-orbital models, respectively. 
Since the Fermi surface is different between the two models for some 
$k_z$ portions, the gap function  differs 
between the two at those portions of the Fermi surface. 
Nonetheless, the present comparison shows that 
the five-orbital model can be considered as reliable  as far as 
the form of the superconducting gap on most of the $k_z$ planes are concerned.
Hereafter, we concentrate on the five-orbital model and 
go down to lower temperatures, where larger-number $k$-point meshes and 
Matsubara frequencies are required.
\section{Five-orbital model at lower temperature}
\subsection{Spin fluctuation}
\begin{figure}[t]
\begin{center}
\includegraphics[width=8cm,clip]{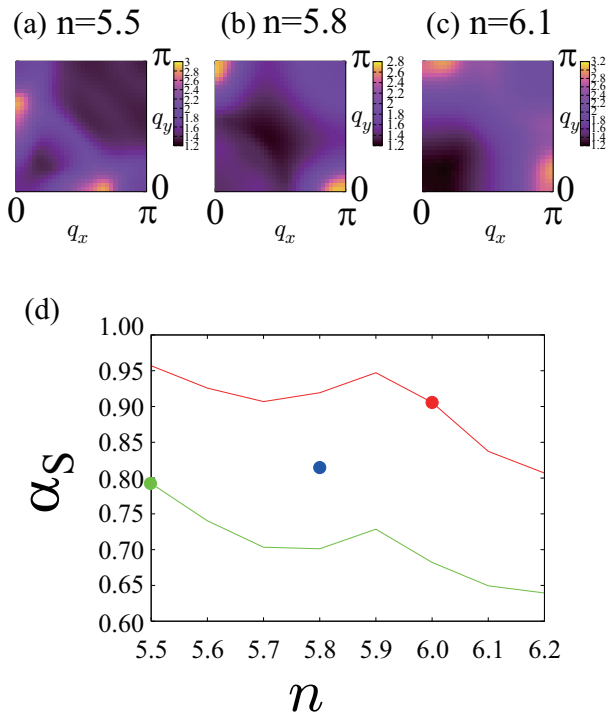}
\end{center}
\caption{(Color online) The spin susceptibility for the five-orbital models 
at $T=0.04$eV with (a) $n=5.5$ (KFe$_2$As$_2$), 
(b) $n=5.8$ (Ba$_{0.6}$K$_{0.4}$Fe$_2$As$_2$), and 
(c) $n=6.1$ (Ba(Fe$_{0.9}$Co$_{0.1}$)$_{2}$As$_2$). 
(d) The Stoner factor against the band filling for the models of 
KFe$_2$As$_2$ and BaFe$_2$As$_2$. The dots indicate the positions of the 
actual band filling of the corresponding material. 
At $n=5.8$, we adopt the model of Ba$_{0.6}$K$_{0.4}$Fe$_2$As$_2$.}
\label{chi}
\end{figure}
\begin{figure}[t]
\begin{center}
\includegraphics[width=8cm,clip]{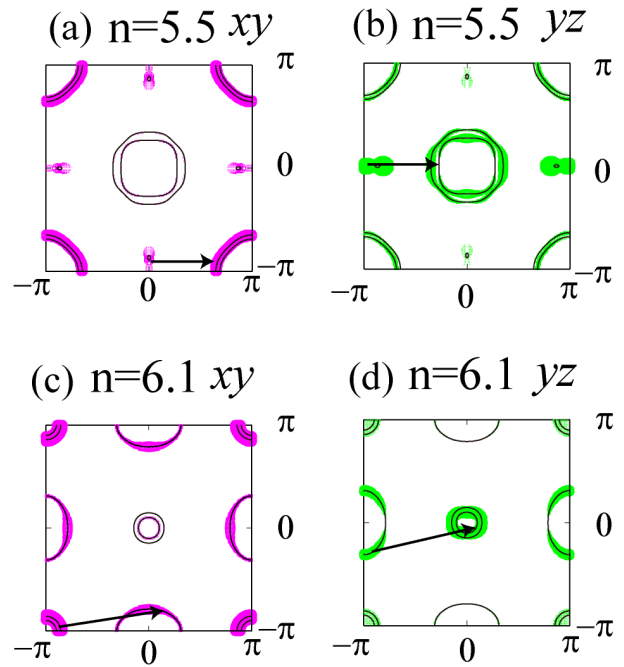}
\end{center}
\caption{(Color online) 
The orbital character distribution on the Fermi surface for $k_z=0$.
$xy$ (a),(c) and $yz$ (b),(d) orbitals for the band fillings 
$n=5.5$ (a),(b) and $n=6.1$ (c),(d). The arrows indicate the 
wave vectors connecting the portions of the Fermi surface 
having similar orbital character. These arrows correspond to the 
peak position in the spin susceptibility.}
\label{orbcha}
\end{figure}
We now discuss the spin fluctuations using the five-orbital model.
Here we take $T=0.04$eV, adopting $64\times 64\times 16$ $k$-point meshes 
and 512 Matsubara frequencies. 
One of the merits to adopting the five-orbital model is that it captures 
the essence of the spin fluctuations observed in neutron-scattering experiments. 
In Fig.\ref{chi}, we show the contour plot of the spin susceptibility 
of the five-orbital model in the unfolded Brillouin zone at $k_z=0$. 
In addition to the results for KFe$_2$As$_2$ (band filling $n=5.5$), 
we also show for comparison the results for Ba$_{0.6}$K$_{0.4}$Fe$_2$As$_2$ ($n=5.8$) 
and Ba(Fe$_{0.9}$Co$_{0.1}$)$_2$As$_2$ $(n=6.1)$.
For Ba$_{0.6}$K$_{0.4}$Fe$_2$As$_2$, we use the five-orbital model that 
is obtained by making a linear combination of the hopping integrals of BaFe$_2$As$_2$ 
and KFe$_2$As$_2$ using the lattice structure of Ba$_{0.6}$K$_{0.4}$Fe$_2$As$_2$ given 
in Ref. \onlinecite{Rotter2}, as was done in Ref. \onlinecite{Suzuki122}.
For $n=6.1$, we simply use the five-orbital model of BaFe$_2$As$_2$ adopting the 
experimental lattice structure.\cite{Rotter1}
The interaction reducing rate is taken to be $f=0.40$, so that the Stoner factor 
does not exceed unity in all of the three cases considered here. 
In Fig. \ref{chi}(a), the peak position is located at an incommensurate 
wave vector [(1-2$\delta$)$\pi$,0], $\delta=0.17$, which is in very good agreement 
with a recent neutron-scattering experiment for  KFe$_2$As$_2$\cite{LeeK122}.

For Ba$_{0.6}$K$_{0.4}$Fe$_2$As$_2$ ($n=5.8$), 
the peak moves to the commensurate position $(\pi,0)$. 
In the case of electron doping, $n=6.1$, the peak moves toward $(\pi,\pi)$ in the 
unfolded Brillouin zone, which is consistent with the experimental 
results of Ba(Fe,Co)$_2$As$_2$ in the high-energy regime. 
The above tendency has also been obtained in other theoretical studies that use 
realistic band structures.\cite{Graser122,Yaresko,Knolle,Kuroki2008,IkedaFLEX2,Park2010}

In the above, the adopted band structure 
is different among Figs. \ref{chi}(a),5(b) and 5(c). 
To focus on the evolution of the spin fluctuations, 
we have also fixed the band structure to that of KFe$_2$As$_2$ or 
BaFe$_2$As$_2$, and varied the band filling hypothetically from 5.5 to 6.2. 
We find that for both of the band structures, 
the spin fluctuation becomes commensurate around $n=5.8\sim 6.0$.
For both of the models, it is found that the peak position continuously moves 
from the commensurate to the incommensurate position as holes are doped,\cite{LT26proc} 
indicating that the spin fluctuation has the same origin 
in the entire (Ba,K)Fe$_2$As$_2$ system, namely, the interband scattering between 
the band around $(0,0)$ and $(\pi,0)/(0,\pi)$. 

The Stoner factor obtained for the above models is shown 
in Fig. \ref{chi}(d) as functions of the band filling. 
This shows that the strength of the spin fluctuations is stronger for the 
model of BaFe$_2$As$_2$ than in KFe$_2$As$_2$ when compared at the 
same band filling. Therefore, when the composition of the 
material is changed continuously from Ba to K, we would expect the 
spin fluctuation to be maximum around the undoped regime (the Ba-rich regime), 
and decreases when the holes are doped by Ba$\rightarrow$K replacement. 
However, in the heavily hole doped regime (the K-rich regime), the strength of the 
spin fluctuations should barely decrease, and may become even 
stronger with hole doping approaching the end material KFe$_2$As$_2$.

As for the electron-hole asymmetry of the incommensurability of the spin fluctuations, 
we have analyzed its origin in Ref. \onlinecite{LeeK122} as follows. 
The spin fluctuations are mainly governed by the interaction 
between portions of the Fermi surface having similar orbital character, 
so in Fig. \ref{orbcha}(a)-(d), we show the strength of the 
orbital character on the Fermi surfaces of 
KFe$_2$As$_2$ $(n=5.5)$ and BaFe$_2$As$_2$ with $n=6.1$ 
in the unfolded Brillouin zone.  
Here we plot the states within a finite energy range $-\Delta E< E(k)-E_F <\Delta E$ 
with $\Delta E=0.02$eV, which can contribute to the spin fluctuations.  
Although Fermi surfaces near the wave vector $(\pi,0)/(0,\pi)$ are barely present 
in KFe$_2$As$_2$, there are states in the vicinity of the Fermi level 
originating from a nearly flat band lying close to the Fermi level.
The position of this flat band with respect to the Fermi level can be clearly seen 
in the magnifications of Figs. \ref{fig1}(a) and 1(b).
As mentioned in Ref. \onlinecite{Kuroki_prb}, the spin fluctuations 
develop at wave vectors which bridge the portions of the 
Fermi surface having similar orbital character. 
The wave vectors connecting $xy$ orbital portions and those connecting 
$yz$ (or $xz$, not shown) portions nearly coincide in each of the two cases, 
so that this wave vector should correspond to the peak position of the spin fluctuations. 
In the electron-doped case, the connecting wave vector deviates 
from $(\pi,0)$ toward $(\pi,\pi)$, while in the hole-doped case it deviates toward $(0,0)$.
Thus, the difference of the spin fluctuation incommensurability 
between electron- and hole-doped cases originates from the multiorbital nature of the system.

\subsection{Superconducting gap}
\begin{figure}[p]
\begin{center}
\includegraphics[width=4.5cm,clip]{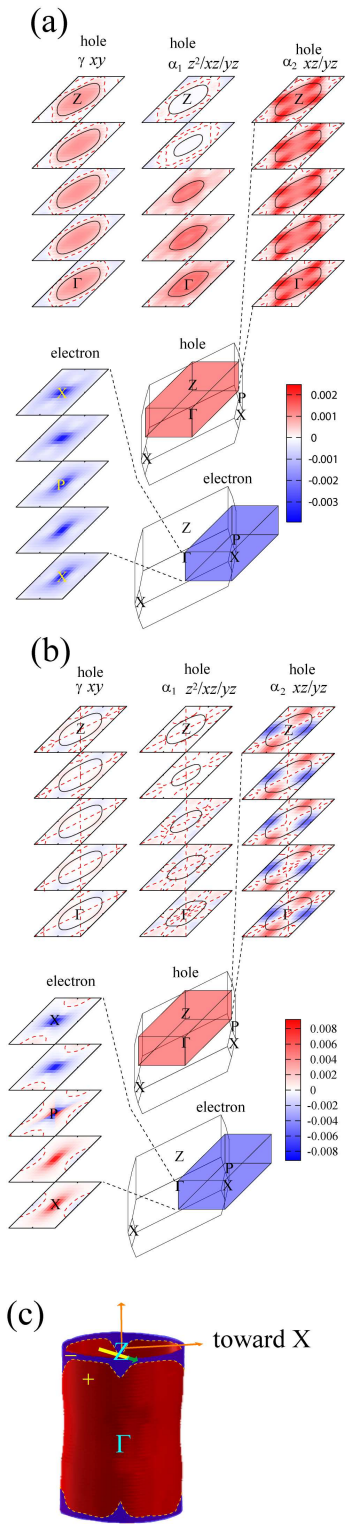}
\end{center}
\caption{(Color online) The contour plots of gap function of the 
five-orbital model for $s$-wave (a) and $d_{x^{2}-y^{2}}$-wave (b) 
at $T=0.04$eV. The solid black lines represent the Fermi surfaces, 
while dashed red lines are the nodes of the gap.
(c) A schematic figure of the horizontal nodes in the $\alpha_1$ Fermi surface 
in the case of $s$-wave pairing. The arrow indicates the 
intraband scattering that drives the sign change.}
\label{fig7}
\end{figure}

Finally, we discuss the superconducting gap of KFe$_2$As$_2$.
We obtain the superconducting gap at $T=0.04$eV 
for $s$-wave and $d_{x^{2}-y^{2}}$-wave pairings.
Here we take $32\times 32\times 16$ $k$-point meshes and 1024
Matsubara frequencies.
We show in Fig. \ref{fig7} 
the $s$-wave (a) and $d$-wave (b) gap functions of the five-orbital model 
refolded into the original folded Brillouin zone. 
Here we take the interaction reducing rate $f=0.49$. 
The rate $f$ is increased compared to the values adopted for 
the spin susceptibility because here we focus only on 
KFe$_2$As$_2$, so that the Stoner factor does not exceed unity 
even for this $f$. We take larger values of the interaction 
so as to enhance the eigenvalue of the Eliashberg equation, 
which enables us to obtain a more reliable form of the superconducting gap.
The eigenvalue of the Eliashberg equation is $\lambda_{s}\sim 1.01$ and 
$\lambda_{d_{x^{2}-y^{2}}}\sim 0.98$, close to unity. 
These values indicate that the pairing symmetry competition is close.
In fact, the possibility of $d$-wave pairing 
was proposed in a recent functional renormalization 
group study\cite{ThomaleK122} 
and an effective Hamiltonian approach.\cite{Maiti}
Even at this low temperature, the gap is essentially similar 
to the one obtained at higher temperature in Sec. \ref{1025sec}.
The $s$-wave gap is basically $s\pm$-wave in the sense that the 
gap function changes sign between the states near $\Vec{k}\sim (0,0)$ and 
$\Vec{k}\sim (\pi,0)/(0,\pi)$. The origin of this sign change is 
strongly related to the spin fluctuation discussed in the present section. 
Namely, although the Fermi surface is barely present around 
$\Vec{k}\sim (\pi,0)/(0,\pi)$, there are states in the vicinity of the 
Fermi level originating from the bands that produces the electron Fermi surface 
for smaller hole content. The pair scattering from those states 
to the hole Fermi surface and vice versa mediated by the incommensurate 
spin fluctuations is the origin of the sign change of the 
gap function. 
On the other hand, the $d$-wave pairing originates essentially due 
to the intraband repulsive interaction. In fact, we find 
that the spin fluctuations originating from the intraband interaction 
(within the hole Fermi surfaces) develops especially for the band filling below $n=5.5$.

Although the origin of the stabilization of the $s$-wave gap is 
the incommensurate spin fluctuation and the gap function 
(eigenfunction of the Eliashberg equation, to be strict) 
changes its sign between $\Vec{k}\sim(0,0)$ and $\Vec{k}\sim(\pi,0)$, 
theoretically, the Fermi surface around the latter wave vector is barely 
present for KFe$_2$As$_2$, so, experimentally, this sign change should 
not be detectable. However, there are several possibilities of (nearly) 
vanishing gap function, 
which may be relevant to the experimental observations that suggest 
the presence of nodes in the superconducting gap.
First, the $s$-wave superconducting gap on the $xy$ 
Fermi surface is rather small. One can see in Fig. \ref{fig7}(a) 
that the nodal line runs somewhat close to the $xy$ Fermi surface.
This may experimentally be detected as nearly vanishing gap, 
although there is no actual sign change in the gap function here.
Second, the $s$-wave gap 
on the $\alpha_1$ Fermi surface has different signs between 
the $k_z$ plane that includes the $\Gamma$ point and 
that with the Z point, as seen in Fig. \ref{fig7}(a). 
This means that the gap has horizontal 
nodes on the $\alpha_1$ Fermi surface, which is schematically 
shown in Fig. \ref{fig7}(c). 
Although the five-orbital model is not reliable concerning 
this portion of the Fermi surface, this sign change is also found in 
the ten-orbital model (at higher temperature) as seen 
in Fig. \ref{1025gap} in Sec.\ref{1025sec}.
The sign change in this $\alpha_1$ hole Fermi surface is 
a common feature found in the RPA studies of 
the 122 systems,\cite{Graser122,Suzuki122}
and indeed some recent experiments\cite{Mukuda,DLFeng} support this 
possibility in phosphorous doped 122 materials.
Third, there is a possibility of a $d$-wave gap, 
which is in close competition with the $s$-wave. 
This possibility, however, may not be consistent with an experiment 
on the magnetic vortex lattice, which shows 
that the gap is not strongly anisotropic within the planes.\cite{Furukawa}
\section{Conclusions}
In the present study, we have constructed an approximate five-orbital 
model of KFe$_2$As$_2$ and compared it with the original 
ten-orbital model. 
We find the former to be reliable as far as the peak position of 
the spin fluctuations and form of the gap function are concerned.
Using the five-orbital model, we have investigated 
the spin fluctuations and the spin-fluctuation-mediated 
superconducting gap of KFe$_2$As$_2$.
We find that the spin fluctuation has an incommensurate peak structure, 
whose origin is the multiorbital nature of the Fermi surfaces. 
The peak position of the spin fluctuation is in excellent agreement with the
experiment.\cite{LeeK122}

As for the superconductivity, $s$-wave and $d$-wave pairings are found to 
be in close competition. The $s$-wave gap has essentially the 
$s\pm$-wave form in the sense that the gap function changes its sign 
between the wave vectors around $(0,0)$ and $(\pi,0)/(0,\pi)$. 
Although there are barely Fermi surfaces around $(\pi,0)/(0,\pi)$, 
the states in the vicinity of the Fermi level are nonetheless 
effective to give the incommensurate spin fluctuations and thus 
the sign change in the superconducting gap function.
Although the sign change exists in the gap function, this should not be 
detectable in experiments because the Fermi surface is barely 
present around $(0,\pi)/(\pi,0)$. As for the explanation for the 
nodes in the superconducting gap observed experimentally, besides a 
possible $d$-wave pairing, we raise two possibilities for the $s$-wave pairing: 
either  horizontal nodes in the $xz/yz/z^2$ Fermi surface 
or the small gap on the entire $xy$ Fermi surface.

\section{acknowledgments}
We are grateful to 
C.H. Lee, H. Fukazawa, H. Ikeda, 
T. Shibauchi Y. Matsuda, H. Eisaki, K. Yamada, 
T. Shimojima, and K. Okazaki for valuable discussions.
Numerical calculations were performed at the facilities of
the Information Technology Center, University of Tokyo, 
and also at the Supercomputer Center,
ISSP, University of Tokyo. 
This study has been partially supported by 
Grant-in-Aid for Scientific Research from  MEXT of Japan and from 
the Japan Society for the Promotion of Science.
%


\begin{thebibliography}{99}
\bibitem{Fletcher} J. D. Fletcher 
, A. Serafin, L. Malone, J. G. Analytis, 
J-H Chu, A.S. Erickson, I. R. Fisher, and 
A. Carrington, 
Phys. Rev. Lett. {\bf 102}, 147001 (2009).
\bibitem{Hicks} C. W. Hicks 
, T. M. Lippman, M. E. Huber, J. G. Analytis, J.-H. Chu, 
A. S. Erickson, I. R. Fisher, and K. A. Moler, 
Phys. Rev. Lett. {\bf 103}, 127003 (2009).
\bibitem{Yamashita} M. Yamashita 
, N. Nakata, Y. Senshu, S. Tonegawa, K. Ikada, 
K. Hashimoto, H. Sugawara, T. Shibauchi, and Y. Matsuda, 
Phys. Rev. B {\bf 80}, 220509(R) (2009).
\bibitem{Hashimoto} K. Hashimoto 
, M. Yamashita, S. Kasahara, 
Y. Senshu, N. Nakata, S. Tonegawa, K. Ikada, A. Serafin, 
A. Carrington, T. Terashima, H. Ikeda, T. Shibauchi and 
Y. Matsuda, Phys. Rev. B {\bf 81}, 220501(R) (2010).
\bibitem{Ishida} Y. Nakai 
, T. Iye, S. Kitagawa, K. Ishida, 
S. Kasahara, T. Shibauchi, Y. Matsuda and T. Terashima: 
Phys. Rev. B. {\bf 81} (2010) 020503(R).
\bibitem{Yamashita2} M. Yamashita, Y. Senshu, T. Shibauchi, 
S. Kasahara, K. Hashimoto, D. Watanabe, H. Ikeda, T. Terashima, 
I. Vekhter, A.B. Vorontsov, and Y. Matsuda,
Phys. Rev. B {\bf 84} 060507(R) (2011).
\bibitem{HashimotoLi111} K. Hashimoto, S. Kasahara, R. Katsumata, 
Y. Mizukami, M. Yamashita, H. Ikeda, T. Terashima, A. Carrington, 
Y. Matsuda, and T. Shibauchi, arXiv:1107.4505.
\bibitem{Mazin122} I. I. Mazin 
, T. P. Devereaux, J. G. Analytis, 
Jiun-Haw Chu, I. R. Fisher, B. Muschler and R. Hackl, 
Phys. Rev. B {\bf 82}, 180502 (2010).
\bibitem{Martin} C. Martin 
, H. Kim, R. T. Gordon, N. Ni, V. G.Kogan, S. L. Bud'ko, 
P. C. Canfield, M. A. Tanatar, R. Prozorov,
Phys. Rev. B {\bf 81}, 060505 (2010).
\bibitem{Kuroki_prb} K. Kuroki 
, H. Usui, S. Onari, R. Arita and H. Aoki, 
Phys. Rev. B {\bf 79}, 224511 (2009).
\bibitem{DHLeeP} F. Wang, H. Zhai, and D.-H. Lee, 
Phys. Rev. B {\bf 81}, 184512 (2010).
\bibitem{IkedaFLEX2} H. Ikeda, R. Arita, and J. Kune\v{s}, Phys. Rev. B 
{\bf 81}, 054502 (2010).
\bibitem{Kariyadoreal} T. Kariyado and M. Ogata, J. Phys. Soc. Jpn. 
{\bf 79}, 033703 (2010).
\bibitem{ThomaleLaFePO} R. Thomale, C. Platt, W. Hanke, B.A. Bernevig, 
Phys. Rev. Lett. {\bf 106}, 187003 (2011).
\bibitem{Sato} T. Sato 
, K. Nakayama, Y. Sekiba, P. Richard, 
Y.-M. Xu, S. Souma, T. Takahashi, G. F. Chen, J. L. Luo, 
N. L. Wang and  H. Ding, 
Phys. Rev. Lett. {\bf 103}, 047002 (2009).
\bibitem{Terashima} T. Terashima 
, M. Kimata, N. Kurita, H. Satsukawa, A. Harada, 
K. Hazama, M. Imai, A. Sato, K. Kihou, C.-H. Lee,
H. Kito, H. Eisaki, A. Iyo, T. Saito, H. Fukazawa, 
Y. Kohori, H. Harima, and  S. Uji, 
J. Phys. Soc. Jpn. {\bf 79}, 053702 (2010).
\bibitem{Fukazawa} H. Fukazawa 
, Y. Yamada, K. Kondo, T. Saito, Y. Kohori, K. Kuga, 
Y. Matsumoto, S. Nakatsuji, H.Kito, P.M. Shirage, K. Kihou, 
N. Takeshita, C. H. Lee, A. Iyo and H. Eisaki, 
J. Phys. Soc. Jpn. {\bf 78}, 083712 (2009).
\bibitem{Dong} J. K. Dong 
, S. Y. Zhou, T. Y. Guan, H. Zhang, Y. F. Dai, X. Qiu, 
X. F. Wang, Y. He, X. H. Chen and S. Y. Li, 
Phys. Rev. Lett. {\bf 104}, 087005 (2010).
\bibitem{Zhang} S.W. Zhang 
, L. Ma, Y.D. Hou, J. Zhang, T.-L. Xia, G.F.Chen, 
J.P. Hu, G.M. Luke and W. Yu, 
Phys. Rev. B {\bf 81}, 012503 (2010).
\bibitem{Matsuda2} K. Hashimoto 
, A. Serafin, S. Tonegawa, R. Katsumata, 
R. Okazaki, T. Saito, H. Fukazawa, Y. Kohori, K. Kihou, 
C. H. Lee, A. Iyo, H. Eisaki, H. Ikeda, Y. Matsuda, A. Carrington 
and T. Shibauchi, Phys. Rev. B {\bf 82}, 014526 (2010). 
\bibitem{ThomaleK122} R. Thomale, C. Platt, W. Hanke, J. Hu,
and B. A. Bernevig, Phys. Rev. Lett. {\bf 107}, 117001 (2011).
\bibitem{Maiti} S. Maiti, M. Korshunov, T. Maier, P.J. Hirschfeld, and 
A.V. Chubukov, Phys. Rev. Lett. {\bf 107}, 147002 (2011), arXiv:1109.0498.
\bibitem{LeeK122} C. H. Lee, 
 K. Kihou, H. Kawano-Furukawa, T. Saito, A. Iyo, 
H. Eisaki, H. Fukazawa, Y. Kohori, K. Suzuki, H. Usui, 
K. Kuroki and K. Yamada, 
Phys. Rev. Lett. {\bf 106},067003 (2011).
\bibitem{struct} S. Rozsa, H. U. Schuster, 
Z. Naturforsch. B {\bf 36}, 1668 (1981).
\bibitem{pwscf} 
S. Baroni 
, A. Dal Corso, S. de Gironcoli, P. Giannozzi, C. Cavazzoni,
G. Ballabio, S. Scandolo, G. Chiarotti, P. Focher, A. Pasquarello,
K. Laasonen, A. Trave, R. Car, N. Marzari, A. Kokalj: 
http://www.pwscf.org/.
Here we adopt the exchange correlation functional introduced by
J. P. Perdew, K. Burke, and Y. Wang
(Phys. Rev. B {\bf 54}, 16533 (1996)), and the wave functions are 
expanded by plane waves up to a cutoff energy of 40 Ry.
8$^3$ $k$-point meshes are used.
\bibitem{MaxLoc} N. Marzari and D. Vanderbilt,
Phys. Rev. B {\bf 56} 12847 (1997); 
I. Souza, N. Marzari and D. Vanderbilt, 
Phys. Rev. B {\bf 65}, 035109 (2001).
The Wannier functions are generated by the code developed by
A. A. Mostofi, J. R. Yates, N. Marzari, I. Souza and D. Vanderbilt,
(http://www.wannier.org/) 
for the energy window $-2.2$ eV $<\epsilon_k-E_F<$ 3.6eV,
where $\epsilon_k$ is the eigenenergy of the Bloch states
and $E_F$ the Fermi energy.
\bibitem{VESTA} K. Momma and F. Izumi,
 J. Appl. Crystallogr. {\bf 41}, 653 (2008).
\bibitem{Kuroki2008}  K. Kuroki 
, S. Onari, R. Arita, H. Usui, Y. Tanaka, 
H. Kontani, and H. Aoki, 
Phys. Rev. Lett. {\bf 101}, 087004 (2008).
\bibitem{GraserComment} A five orbital model for 
BaFe$_2$As$_2$ was constructed in ref.\onlinecite{Graser122}, 
which has a different orbital character in the $\alpha_1$ Fermi surface
compared to our model.
\bibitem{Miyake} T. Miyake 
, K. Nakamura, R. Arita, and M. Imada,  
J. Phys. Soc. Jpn. {\bf 79}, 044705 (2010).
\bibitem{Rotter2} M. Rotter, 
 M. Tegel, D. Johrendt, 
Phys. Rev. Lett. {\bf 101}, 107006 (2008).
\bibitem{Suzuki122} K. Suzuki, H. Usui, K. Kuroki ,
J. Phys. Soc. Jpn. {\bf 80}, 013710 (2010)
\bibitem{Rotter1} M. Rotter, 
M. Tegel, D. Johrendt, I. Schellenberg,
 W. Hermes and R. P\"ottgen, Phys. Rev. B {\bf 78}, 020503 (2008).
\bibitem{Graser122} S. Graser, A. F. Kemper, T. A. Maier, 
H.-P. Cheng, P. J. Hirschfeld, and D. J. Scalapino, 
Phys. Rev. B {\bf 81}, 214503 (2010).
\bibitem{Park2010} J. T. Park 
, D. S. Inosov,A. Yaresko, S. Graser, D. L. Sun, Ph. Bourges, 
Y. Sidis, Y. Li, J.-H. Kim, D. Haug, A. Ivanov, K. Hradil,
A. Schneidewind, P. Link, E. Faulhaber, I. Glavatskyy, 
C. T. Lin, B. Keimer, and V. Hinkov,
Phys. Rev. B. {\bf 82}, 134503 (2010).
\bibitem{Knolle} J. Knolle, I. Eremin, and R. Moessner
Phys. Rev. B {\bf 83}, 224503 (2011)
\bibitem{Yaresko} A. N. Yaresko, G.-Q. Liu, V. N. Antonov, and 
O. K. Andersen,  Phys. Rev. B {\bf 79}, 144421 (2009).
\bibitem{LT26proc} K. Suzuki, H. Usui, and K. Kuroki,  
to be published in the proceedings of LT26
\bibitem{Mukuda}
T. Dulguun, H. Mukuda, H. Kinouchi, M. Yashima, Y. Kitaoka,
T. Kobayashi, S. Miyasaka, and S. Tajima, arXiv:1108.4480.
\bibitem{DLFeng}
Y. Zhang, Z. R. Ye, Q. Q. Ge, F. Chen, J. Jiang,
M. Xu, B. P. Xie, and  D. L. Feng, arXiv:1109.0229.
\bibitem{Furukawa} H. Kawano-Furukawa 
, C. J. Bowell, J. S. White, R.W. Heslop, A.S. Cameron,
E.M. Forgan, K. Kihou, C. H. Lee, A. Iyo, H. Eisaki,
T. Saito, H. Fukazawa, Y. Kohori, R. Cubitt,
C. D. Dewhurst, J. L. Gavilano and M. Zolliker, 
Phys. Rev. B {\bf 84}, 024507 (2011).
\end{thebibliography}
\end{document}